\documentclass[pra,twocolumn,amsmath,eqsecnum,amssymb,superscriptaddress,nofootinbib,showpacs]{revtex4-1}
\usepackage{graphicx}
\usepackage{color}

\usepackage{amsmath}
\usepackage{amssymb}
\usepackage{bm}
\usepackage{float}

\begin{document}

\title{Multiband $d-p$ model and self-doping in the electronic structure of Ba$_2$IrO$_4$}
\thanks{This work is dedicated to Professor J\'ozef Spa\l{}ek 
on the occasion of his 70$^{th}$ birthday.}

\author{     Krzysztof Ro\'sciszewski}
\affiliation{Marian Smoluchowski Institute of Physics, Jagiellonian University,
             prof. S. \L{}ojasiewicza 11, PL-30348 Krak\'ow, Poland }

\author{     Andrzej M. Ole\'s  }
\affiliation{Marian Smoluchowski Institute of Physics, Jagiellonian University,
             prof. S. \L{}ojasiewicza 11, PL-30348 Krak\'ow, Poland }
\affiliation{Max-Planck-Institut f\"ur Festk\"orperforschung,
             Heisenbergstrasse 1, D-70569 Stuttgart, Germany }

\date{\today}

\begin{abstract}
We introduce and investigate the multiband $d-p$ model describing a 
IrO$_4$ layer (such as realized in Ba$_2$IrO$_4$) where all $34$ 
orbitals per unit cell are partly occupied, i.e., $t_{2g}$ and $e_g$ 
orbitals at iridium and $2p$ orbitals at oxygen ions. The model takes 
into account anisotropic iridium-oxygen $d-p$ and oxygen-oxygen $p-p$ 
hopping processes, crystal-field splittings, spin-orbit coupling, and 
the on-site Coulomb interactions, both at iridium and at oxygen ions. 
We show that the predictions based on assumed idealized ionic 
configuration (with $n_0=5+4\times 6=29$ electrons per IrO$_4$ unit) 
do not explain well the independent \textit{ab initio} data and the 
experimental data for Ba$_2$IrO$_4$. Instead we find that the total 
electron density in the $d-p$ states is smaller, $n=29-x<n_0$ ($x>0$). 
When we fix $x=1$, the predictions for the $d-p$ model become more 
realistic and weakly insulating antiferromagnetic ground state with 
the moments lying within IrO$_2$ planes along (110) direction is found, 
in agreement with experiment and \textit{ab initio} data. We also show 
that:
(i) holes delocalize over the oxygen orbitals and the electron density 
at iridium ions is enhanced, hence  
(ii) their $e_g$ orbitals are occupied by more than one electron and 
have to be included in the multiband $d-p$ model describing iridates. 
\end{abstract}

\pacs{71.10.Fd, 71.70.Ej, 74.70.Pq, 75.10.Lp}

\maketitle

\section{Introduction}
\label{intro}

Iridates such as Sr$_2$IrO$_4$ and Ba$_2$IrO$_4$ are transition metal 
oxides with party occupied, spatially extended $5d$ orbitals. They 
belong to strongly correlated systems and are weak antiferromagnetic 
(AF) insulators, with a complex competition of local Coulomb 
interactions, Hund's exchange coupling, crystal-field effects and very 
strong spin-orbit interactions. The debate continues about whether 
these systems are better viewed as a realization of:
(i) spin-orbit Mott insulators \cite{Moo09}, or
(ii) an old idea of a Slater insulator with insulating behavior 
resulting from long-range AF order in a correlated electronic band 
\cite{Car91} is also at work in iridates \cite{Ari12,li13}, or finally  
(iii) a mixture of both above scenarios interrelated with each other 
and contributing almost equally~\cite{Hsi12}. 

The electronic structure of Sr$_2$IrO$_4$ (and to a lesser extent of
Ba$_2$IrO$_4$) was a subject of numerous investigations
\cite{Ari12,Moo06,Wat10,kim12} on a different level of sophistication.
The interest in Sr$_2$IrO$_4$ is motivated by its structural 
similarity to cuprates \cite{Kimx2}, and indeed $d$-wave 
superconductivity was predicted in electron doped Sr$_2$IrO$_4$ by 
numerical studies \cite{Wat13}. Recently mapping of the three-band 
($d-p$) model to a single band was presented and it was shown that 
further neighbor hopping is necessary to describe the difference between 
hole and electron doped iridates \cite{Ham15}. The recent experimental 
evidence seems to support the scenario that Ba$_2$IrO$_4$ is a close 
realization of spin-orbit Mott insulator rather than a Slater system 
\cite{Oka11,Mos14}, similar to Sr$_2$IrO$_4$ investigated using 
angle-resolved photoemission \cite{Tor15}, optical conductivity, x-ray 
absorption measurements, and first-principles band calculations 
\cite{Kim08}. Recently it has been shown that strong spin-orbit 
coupling changes radically the electronic states in Mott insulators 
\cite{Jac09,Brz15}. Within this limit of an insulator, strong spin-orbit 
interaction accompanied by large crystal-field effects split $t_{2g}$ 
orbitals of Ir$^{4+}$ ions into fully filled manifold with effective 
total angular momentum $J_{\rm eff}=3/2$ and singly occupied manifold 
$J_{\rm eff}=1/2$ (half-filled ground state) \cite{Bos13,Sal14}. 
Consequently, tight-binding model calculations were performed using a 
model for $t_{2g}$ orbitals which includes the above effects. It gives 
results in agreement with the angle-resolved photoemission data for 
the occupied electronic bands \cite{Mos14}. 

The origin of AF order in the ground state is more subtle. 
It was shown by \textit{ab initio} calculations that the Heisenberg 
superexchange is the largest low-energy scale but also Ising-like 
compass interactions contribute \cite{Kat14}. Therefore, individual 
Ba$_2$IrO$_4$ layers provide a close realization of the quantum 
spin-1/2 compass-Heisenberg model \cite{Tro10}. We note that similarly 
complex structure of the superexchange was established for the 
honeycomb lattice compound Na$_2$TiO$_3$, where it takes the form of 
Kitaev-Heisenberg model which is under intense discussion at present 
\cite{Cha10}.

However, there remain some serious doubts (supported by experimental 
data) whether such a simplified scenario focused on $t_{2g}$ orbitals 
is indeed fully realistic \cite{Uch14}. One has to admit that a simple 
picture and a model as simple as possible are useful tools when it 
comes to interpretation of rather complicated experimental data. 
However, this is only one level of physical description. On the second 
level one requires much better understanding of what is going on. This 
is the aim of the present paper --- we wish to verify to what extent the 
charge transfer from Ba ions to IrO$_4$ units is complete and whether 
the ionic model could be used to describe the electronic structure of 
Ba$_2$IrO$_4$. We are not satisfied with an idealistic picture of 
Ir$^{4+}$ (with $e_g$ orbitals being empty) surrounded by O$^{2-}$ ions 
and we want to investigate the true charge densities and the magnetic 
order parameters realized in this system. To this end we constructed 
a multiband $d-p$ model and performed unrestricted Hartree-Fock (HF) 
computations on a finite IrO$_4$ cluster which contains $4\times 4$ Ir 
ions and the accompanying $4\times 4\times 4$ oxygen ions --- half of 
them located within the same plane as Ir ions, while the second half 
being in out-of-plane (apical) positions.
The model involves (per a single IrO$_4$ unit) five $5d$ orbitals at 
each Ir and $4\times 3$ oxygen $2p$ orbitals per unit cell, and these 
orbitals are occupied by:
(i) $n_0=5+4\times 6$ electrons, according to the formal and
idealized ionic model, or
(ii) a lower electron number by one electron, i.e., $n=4+4\times 6$ 
electrons, according to the realistic \textit{ ab initio} computations 
\cite{Sto12}.

The paper is organized as follows. In Sec. \ref{sec:model} we introduce
the multiband model which includes all $5d$ states at iridium ions and 
$2p$ states at oxygen ions. The parameters of the model are specified 
in Sec. \ref{sec:para}. The Hartree-Fock approximation for the Coulomb 
interactions is explained in Sec. \ref{sec:hf}, while in Sec. 
\ref{sec:resu} we present the results of numerical calculations:
for the formal ionic model in \ref{sec:ionic},
and for the model with self-doping in Sec. \ref{sec:self}. Here  
we also discuss the justification of finite self-doping (with respect 
to the electron densities in the ionic model) and the possible effect 
of electronic correlations. The paper is concluded with a short 
discussion and summary of the main results  in Sec. \ref{sec:summa}.

\section{Model Hamiltonian}
\label{sec:model}

The $d-p$ Hamiltonian for IrO$_4$ plane reads, %consists of several parts,
\begin{equation}
{\cal H}= H_{\rm kin}+ H_{\rm diag}+H_{\rm so} +H_{\rm int}.
\label{model}
\end{equation}
It consists of several parts ---
the different terms in Eq. (\ref{model}) stand for the kinetic energy
($H_{\rm kin}$), spin-orbit coupling ($H_{\rm so}$), crystal-field
splittings  ($H_{\rm diag}$), and the intraatomic Coulomb interactions
($H_{\rm int}$).

In Sec. \ref{sec:resu} we report a study of charge space-homogeneous
solutions. All symmetry-equivalent occupation numbers, i.e., HF 
primary order parameters, are assumed to be the same in different parts 
of the cluster. When studying the possibility of antiferromagnetism 
there are two sublattices and the number of order parameters doubles. 
Looking for charge space-homogeneous ground states can be considered 
to be a consequence of strong long-range inter-ionic electrostatic 
interactions. These interactions are not explicitly included in the 
model (\ref{model}) but to some extent they are accounted for by the
homogeneity assumption.

\subsection{Kinetic energy}

The kinetic part is:
\begin{equation}
H_{\rm kin}= \sum_{ \{i\mu; m\nu\},\sigma} \left(t_{ i\mu; m\nu}
 c^{\dagger}_{i\mu,\sigma}  c_{m\nu,\sigma}^{} + {\rm H.c.}\right),
\end{equation}
where $c_{i\nu,\sigma}^{\dagger}$ stands for the creation of an 
electron at site $i$ in an orbital $\nu$ with spin 
$\sigma=\uparrow,\downarrow$. The model includes all $5d$ orbital states 
$\nu\in\{xy,yz,zx,x^2-y^2,3z^2-r^2\}$ per Ir atom at site $m$ (in this 
order), and three $2p$ orbitals $\nu\in\{p_x,p_y,p_z\}$ per oxygen atom 
at site $i$. When choosing an alternative and more intuitive notation, 
one can write creation operators $d_{m\nu,\sigma}^{\dagger}$ for 
electron creation in $d$ orbitals and $p_{i\nu,\sigma}^{\dagger}$ 
for $p$ orbitals, as used in Sec. \ref{sec:int}.

The matrix elements $t_{i,\mu;m,\nu}$ are assumed to be non-zero only 
for nearest neighbor iridium-oxygen $p-d$ pairs, and a similar formula 
stands for nearest neighbor oxygen-oxygen $p-p$ pairs (i.e., next 
nearest hopping elements are neglected). The geometry of the cluster is 
identical to that for RuO$_4$ layer (as realized in Sr$_2$RuO$_4$), 
thus the matrix elements $t_{i,\mu;j,\nu}$ can be directly adapted from 
the Appendix of Ref. \cite{rosc15}.

\subsection{Crystal-field splittings}
\label{sec:cf}

Let us now present the diagonal part $H_{\rm diag}$ of the $d-p$ 
Hamiltonian (\ref{model}). It depends only on electron number operators,
and takes into account the effects of crystal field and the difference 
of reference orbital energies (here we employ the electron notation),
\begin{equation}
\Delta=\varepsilon_d-\varepsilon_p,
\label{Delta}
\end{equation}
between $d$ and $p$ orbitals, both for empty states (i.e., when 
neglecting the interaction terms from $H_{\rm int}$, see below). 
We remark that in principle $\varepsilon_p$ might be different 
for apical oxygens and for in-plane oxygens. A large difference 
$\sim 1$ eV between these level positions was considered in 
ruthenate perovskites \cite{oguchi95} as a possibility. 
We have found within the present $d-p$ model (also in a ruthenate 
\cite{rosc15}) that such a difference is definitely too large and 
would overestimate the difference between charge densities. 
Qualitatively, the main factor responsible for the charge anisotropy 
between $2p$ orbitals at these nonequivalent oxygen position is their 
weaker hybridization with $5d$ orbitals, while the difference between 
the corresponding level energies is of secondary importance.

We put the reference orbital energy $\varepsilon_d$ for Ir($5d$) states 
to be zero, hence we use only $\varepsilon_p$ as a parameter and write:
\begin{eqnarray}
H_{\rm diag}  &=&
\sum_{i,\mu=x,y,z;\sigma}
\varepsilon_p^{}\, p^\dagger_{i,\mu,\sigma}p_{i,\mu,\sigma}^{}\nonumber \\
&+&\sum_{m,\mu=xy,yz,... ;\sigma}
f^{}_{\mu,\sigma}\, d^\dagger_{m,\mu,\sigma} d_{m,\mu,\sigma}^{}.
\end{eqnarray}
Here the first sum is restricted to oxygen sites, while the second one
runs over iridium sites.
The vector containing the
elements of $\{f^{}_{\mu,\sigma}\}$ is
\begin{equation}
f^{} = \frac{1}{3}\,D_1 \,\left[
\begin{array}{r} 2\\-1\\-1\\0\\0\\2\\-1\\-1\\0\\0 \end{array} \right]
+ D_2 \, \left[
\begin{array}{r} 0\\0\\0\\1\\0\\0\\0\\0\\1\\0 \end{array} \right]
+ D_3 \, \left[
\begin{array}{r} 0\\0\\0\\0\\1\\0\\0\\0\\0\\1 \end{array} \right].
\end{equation}
It includes the orbital splittings of $5d$ orbitals at Ir ions in the 
tetragonal crystal field. The constant $D_1$ serves as a crude estimate
of the splitting between the orbital $xy$ and the orbital doublet 
$\{yz,zx\}$. The $\{D_2,D_3\}$ parameters refer to splitting between
$t_{2g}$ and $e_g$ orbitals at iridium and are much larger than $D_1$,
see below.

Jahn-Teller part will be neglected in the Hamiltonian (\ref{model}).
The exception is the elongation of bonds between iridium and apical
oxygens which could be considered as static and global $Q_3$ 
Jahn-Teller distortion, but it is simpler to include it by a proper 
renormalization of the crystal-field splittings. Note that in 
Sr$_2$IrO$_4$ Jahn-Teller effects are not negligible due to distortions 
of octahedra, while they vanish for Ba$_2$IrO$_4$
(see Fig. 1 in Ref. \cite{Mos14}).

\subsection{Spin-orbit coupling}

Formally, spin-orbit part $H_{\rm so}$ of the Hamiltonian
Eq. (\ref{model}) has similar mathematical structure to the kinetic
part $H_{\rm kin}$ \cite{rosc15,Miz96a,Pol12},
with $t^{so}_{\mu,\sigma;\nu,\sigma'}$ elements restricted to single
iridium sites,
\begin{equation}
H_{\rm so}\!=\!\sum_m  H_{\rm so}^{(m)}
\!=\!\sum_m \left\{ \sum_{  \mu \neq \nu;\sigma, \sigma }\!
t^{\rm so}_{\mu,\sigma;\nu,\sigma'}
d^{\dagger}_{m\mu,\sigma}d_{m\nu,\sigma'}^{}+{\rm H.c.}\right\},
\label{so-part}
\end{equation}
where the summation runs only over iridium sites. The matrix elements 
are all proportional to spin-orbit coupling strength $\zeta$ which is 
large on Ir sites and usually assumed to be about $\zeta=0.4$ eV 
\cite{kim12}.
Just like the hopping 
elements also the elements $t^{so}_{\mu,\sigma;\nu,\sigma'}$ can be 
directly adapted from Ref. \cite{rosc15} (for the explicit entries for 
$t^{so}_{\mu,\sigma;\nu,\sigma'}$ see Eq. (2.4) in Ref. \cite{rosc15}).
As we use the basis of real $5d$ orbitals (and not the spherical
harmonics) several spin-orbit elements turn out to be imaginary
(thus our Hamiltonian (\ref{model}) is complex). Note that the 
consequence of finite spin-orbit coupling is that the total spin of 
the system and total $z$th spin component are not conserved quantities.

\begin{widetext}
\subsection{Local Coulomb interactions}
\label{sec:int}

The last part of the multiband $d-p$ Hamiltonian, 
$H_{\rm int}\equiv H^d_{\rm int}+H^p_{\rm int}$, stands for strong local 
on-site interactions. For the $d$ orbitals at iridium sites it reads,
\begin{eqnarray}
H^d_{\rm int}&=&
  U_d \sum_{m, \mu}  n_{m, \mu, \uparrow} n_{m, \mu, \downarrow}
+\frac{1}{2}\sum_{i,\mu\neq\nu}
\left(U_d-\frac{5}{2}J_{d,\mu\nu}\right)n_{m,\mu}n_{m,\nu} \nonumber\\
&-& \sum_{i,\mu\neq\nu} 
J_{d,\mu\nu}\,\mathbf{S}_{m,\mu}\cdot\mathbf{S}_{m,\nu}
+  \sum_{m,\mu\neq\nu} J_{d,\mu\nu}\,
d^\dagger_{m,\mu, \uparrow} d^\dagger_{m,\mu, \downarrow}
d_{m,\nu, \downarrow}^{} d_{m,\nu, \uparrow}^{}.
\label{hubbard-intra}
\end{eqnarray}
where $J_{d,\mu\nu}$ is the tensor of on-site inter-orbital exchange
elements for $d$ orbitals which can be expressed using Racah parameters
$B$ and $C$ \cite{Ole05,Gri71} (see also Table I given by Horsch in 
Ref. \cite{Hor07}). The anisotropy between different Hund's exchange 
elements $\{J_{d,\mu\nu}\}$ vanishes only for orbitals of the same 
symmetry, i.e., either in pure $t_{2g}$ system or in pure $e_g$ system 
\cite{Ole12}). For convenience, we rewrite Eq. (\ref{hubbard-intra}) 
to separate it into the diagonal terms in electron densities (first 
line) and the quantum fluctuating part (second line) as follows,
\begin{eqnarray}
H^d_{\rm int}&=&
  U_d \sum_{m, \mu}  n_{m\mu, \uparrow} n_{m\mu, \downarrow}
+\frac{1}{2}\sum_{m\mu\neq\nu,\sigma} \left( U_d-3J_{d,\mu\nu}\right)
 n_{m\mu,\sigma} n_{m\nu,\sigma}
 + \frac{1}{2} \sum_{m\mu\neq\nu,\sigma} \left(U_d-2J_{d,\mu\nu}\right)
 n_{m\mu, \sigma}n_{m\nu,-\sigma}   \nonumber\\
&-& \sum_{m,\mu\neq\nu} J_{d,\mu\nu}\,
d^\dagger_{m\mu \uparrow} d_{m\mu, \downarrow}^{}
d^\dagger_{m\nu \downarrow} d_{m\nu, \uparrow}^{}
+  \sum_{i,\mu\neq\nu} J_{d,\mu\nu}\,
d^\dagger_{m\mu, \uparrow} d^\dagger_{m\mu, \downarrow}
d_{m\nu, \downarrow}^{}d_{m\nu, \uparrow}^{}.
\label{hubbard2-intra}
\end{eqnarray}
The formula for local Coulomb interactions $2p$ orbitals at oxygen 
sites, $H^p_{\rm int}$, is analogous,  
\begin{eqnarray}
H^p_{\rm int}&=&
  U_p \sum_{i, \mu}  n_{i\mu, \uparrow} n_{i\mu, \downarrow}
+\frac{1}{2}\left( U_p-3J_{p}\right)\sum_{i,\mu\neq\nu,\sigma} 
 n_{i\mu,\sigma} n_{i\nu,\sigma}
 + \frac{1}{2}\left(U_p-2J_{p}\right) \sum_{i,\mu\neq\nu,\sigma} 
 n_{i\mu, \sigma}n_{i\nu,-\sigma}   \nonumber\\
&-& \sum_{i,\mu\neq\nu} J_{p}\,
p^\dagger_{i\mu,\uparrow} p_{i\mu,\downarrow}^{}
p^\dagger_{i\nu,\downarrow} p_{i\nu,\uparrow}^{}
+  \sum_{i,\mu\neq\nu} J_{p}\,
p^\dagger_{i\mu, \uparrow} p^\dagger_{i\mu, \downarrow}
p_{i\nu, \downarrow}^{}p_{i\nu, \uparrow}^{}.
\label{hub-p}
\end{eqnarray}
\end{widetext}
and is defined by the intraatomic Coulomb element $U_p$ and Hund's 
exchange $J_p$, as the tensor $J_{p,\mu\nu}$ has identical elements
$J_p$ for all off-diagonal pair of the oorbitals of the same symmetry. 
Here $n_{i\mu,\sigma}\equiv p^\dagger_{i\nu,\sigma}p_{i\nu,\sigma}^{}$ 
and $p^\dagger_{i\mu,\sigma}$ operators refer to the density and 
electron creation within $(\mu,\sigma)$ spin-orbital at site $i$.

\section{The Hamiltonian parameters}
\label{sec:para}

\subsection{Previous studies and deducing the parameters}

The effective $d-p$ model (\ref{model}) requires a choice of a number 
of explicitly included parameters. They may be to a large extent 
deduced from the previous \textit{ab initio} studies. We have adopted 
the values of in-plane hopping elements $(pd\sigma)$ and $(pd\pi)$ 
used in Ref. \cite{kim12} for Sr$_2$IrO$_4$, and rescaled them 
(both for in-plane and out-of-plane bonds) using Harrison 
formulas \cite{Har05} to fit the bond lengths in the structure of 
Ba$_2$IrO$_4$ (reported by Moser \textit{et al.} \cite{Mos14}). 
The elements $(pp\sigma)$ and $(pp\pi)$ were taken directly from 
Ref. \cite{rosc15} and were rescaled in a similar way.

The choice of the Coulomb elements is rather difficult due to their
unknown screening. There are many reliable estimations for $U_d$ in 
effective models featuring only $5d$ Wannier orbitals at iridium ions 
($2p$ oxygen orbitals are absent, but effective iridium $5d$ orbitals 
are then renormalized by hybridization with oxygen $2p$ orbitals). No 
need to say that such parameters would have to be completely different 
in the framework of the multiband $d-p$ model. A rather low value of 
$U_d\in[1.0,2.0]$ eV was first suggested by Mazin \textit{et al.} 
\cite{Maz12}, but we argue that a more probable value is $U_d=3.0$ eV. 
This value (and $J_d=0.6$ eV) were used in local density approximation 
(LDA+$U$) computations for Na$_2$IrO$_3$ \cite{Com12,Li15} and also for 
Ba$_2$IrO$_4$ \cite{Mos14}. A similar value of 
$U_d=2.72$ eV was obtained by constrained random phase approximation in 
Na$_2$IrO$_3$ \cite{Yam14}. Other estimates are: $U_d\simeq 1.9$ eV 
\cite{Ari12} and $U_d=2.5$ eV \cite{Uch14,li13}. Following this 
discussion and arguments presented in Ref. \cite{li13}, we decided to 
fix $U_d=2.5$ eV.

Hund's exchange elements are less screened than intraorbital Coulomb
elements and are closer to their atomic values. For Hund's exchange 
$J_d$ between two $t_{2g}$ electrons we selected the value 0.5 eV, in
agreement with the old semiempirical prescription ($J_d/U_d\simeq 0.2$) 
\cite{hunds}. The same value was used in Refs. \cite{kim12,Maz12} while  
slightly smaller ($J_d=0.4$ eV \cite{Mos14}), 
very small ($J_d=0.14$ eV \cite{Ari12}), or larger 
($J_d=0.6$ eV \cite{Com12,Li15}) values were also considered. For the 
sake of fixing precisely Hund's coupling tensor elements $J_{d,\mu\nu}$ 
we use Table I from Ref. \cite{Hor07} and in addition we use a 
semiempirical formula $C\simeq 4B$ for Racah parameters. With this 
\textit{Ansatz} one finds $J_d=3B+C\approx 7B$ for a pure $t_{2g}$ 
system \cite{Ole12}. The $J_{d,\mu\nu}$ elements concerning $e_g$ 
orbitals are finite and fixed again using the entries from 
Table I in Ref. \cite{Hor07}.

\begin{figure}[t!]
\vskip -2.7cm
\begin{center}
\includegraphics[width=14cm]{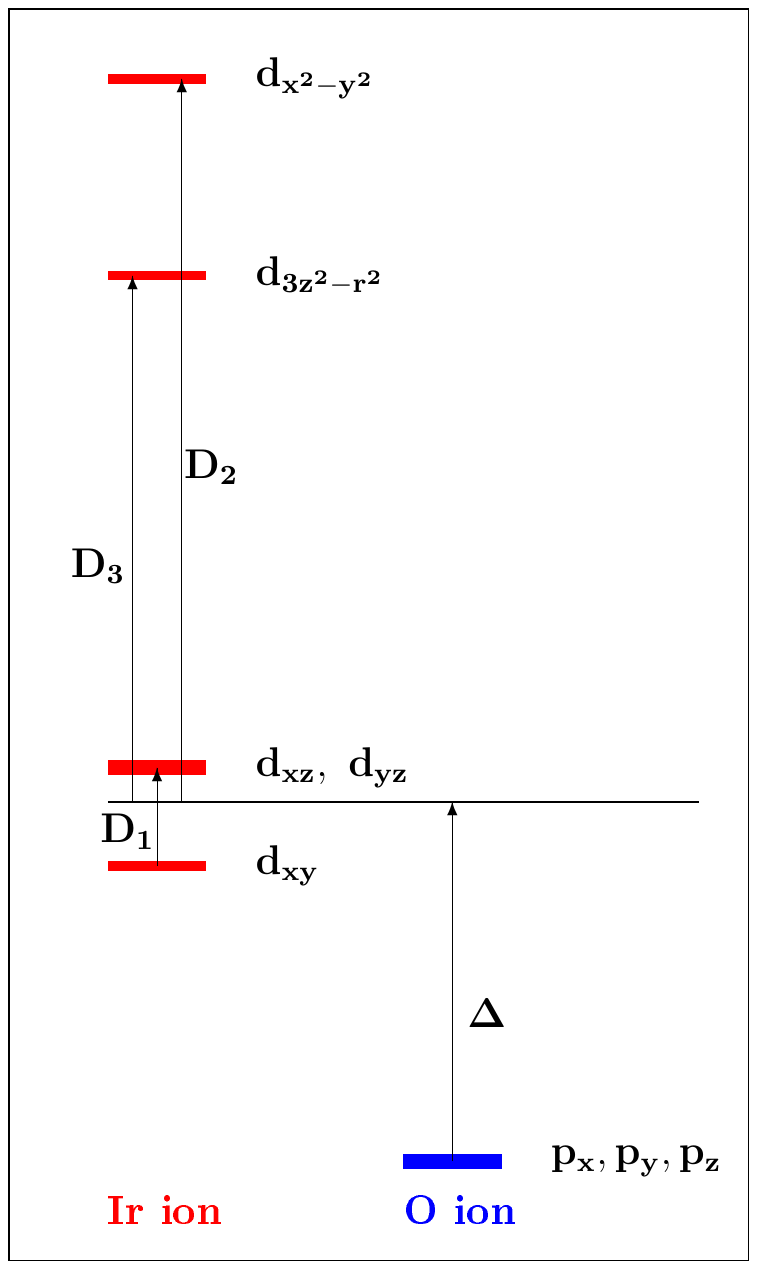}
\end{center}
\vskip -8.2cm
\caption{(Color online)
Artist's view of the $5d$ orbital energies split by the elements 
$\{D_1,D_2,D_3\}$ at Ir ions in the absence of spin-orbit coupling 
($\zeta=0$) (left), and the $d-p$ splitting $\Delta$ Eq. 
(\ref{Delta}) between the Ir($5d$) and O($2p$) orbitals (right).}
\label{fig:ddd}
\end{figure}

Unfortunately, not much is known about the intraorbital Coulomb 
repulsion $U_p$ at oxygen ions (in iridates). In Ref. \cite{Kho03} 
this parameter is estimated to be within $U_p\in[4.0,6.0]$ eV. We 
recall that in cuprates the values $U_p=4.5$ eV \cite{Arr09} and 
$U_p\simeq 4.0$ eV \cite{Hyb89,Gra92} were considered; note that 
several other values were also suggested (all in the range 3-8 eV), 
while 6.0 eV was indicated by some experimental data \cite{Esk91}. We 
use below $U_p=4.4$ eV. For Hund's coupling $J_p$ at oxygen ions the 
values $J_p=0.6-0.8$ eV were suggested \cite{Kho03}, while Grant and 
McMahan computations in cuprates yield $J_p=0.8$ eV \cite{Gra92}. 
Following these estimates we use below as well $J_p=0.8$ eV. 

To complete the set of the Hamiltonian parameters we must provide 
estimates for the crystal-field splittings of Sec. \ref{sec:cf}. 
These are: $D_1=-0.07$ eV, $D_2=2.6$ eV, and  $D_3=2.0$ eV, and we 
adopted them from \textit{ab initio} computations \cite{Sto12}. The 
$5d$ orbital degeneracy is lifted by these terms and the orbital 
states at Ir ions for $\zeta=0$ are shown in Fig. \ref{fig:ddd}. 
The $e_g$ orbitals are much higher than $t_{2g}$ ones, and the 
highest orbital energy is found for $x^2-y^2$ orbital, similar to 
the situation in cuprates for the corresdponding $3d$ orbital 
\cite{Hyb89,Gra92}. 

The charge-transfer gap $\Delta$ (\ref{Delta}) will be examined as a 
continuous parameter in the range from 1.0 eV up to 5.0 eV, and the 
representative situation is shown in Fig. \ref{fig:ddd}. Indeed, oxygen 
$2p$ orbital energies are below the Fermi energy, but hybridization 
with $5d$ orbitals is responsible for partial electron transfer towards 
Ir($5d$) orbitals analyzed below in Sec. \ref{sec:resu}. We have taken 
$\Delta$ as a free parameter as its precise value is unknown. There are 
only a few values for $\Delta$ in the literature, and these data are 
inconclusive. In Refs. \cite{Moo06,kim12} $\Delta=3.3$ eV was used. 
According to \textit{ab initio} computations \cite{Sto12} of 
charge-transfer excitations, the value of $\Delta$ is about 2.0 eV or 
slightly larger --- this value however is not a bare parameter but 
includes spin-orbit coupling plus strong correlation effects; thus the 
bare value could be significantly larger. In Ref. \cite{Shi08} the 
value of $\Delta=2.0$ eV is given for Na$_2$IrO$_3$. 

Finally, the spin-orbit coupling strength $\zeta$ on Ir sites is large 
--- it modifies the orbital states and mixes the spin states at iridium 
ions \cite{Jac09}. It is usually assumed to be 0.4 eV \cite{kim12}. 
A more accurate value is 0.43 eV \cite{Sal14,Sto12} and we use it in 
the present study. 

All the parameters we use in the calculations below are collected in 
Table I. Note that during computations we are setting the value of 
$\varepsilon_d$ to be zero as the reference energy. 
Note that the value of parameter $(pd\sigma)$ which is involved in
hopping processes from $t_{2g}$ to $e_g$ orbitals is $-1.69$ eV, see 
Appendix in Ref. \cite{rosc15}, and the splitting between $t_{2g}$ 
and $e_g$ orbitals is $2.0 - 2.6$ eV. This suggests that the 
expectation that $e_g$ levels are empty is unrealistic.

\begin{table}[t!]
\caption{Parameters of the Hamiltonian (\ref{model}) (all in eV) used 
in the HF calculations. For the hopping integrals we present only 
representative in-plane Slater integrals $(pd\pi)$ and $(pp\pi)$. 
Out-of-plane integrals are obtained by applying Harrison scaling 
\cite{Har05}. The charge-transfer gap, 
$\Delta = \varepsilon_d-\varepsilon_p$, is examined in the range 
$\Delta\in[1.0,5.0]$ eV.
}
\begin{ruledtabular}
\begin{tabular}{cccccc}
  $U_d$ & $J_d$  & $U_p$ & $J_p$ & $\zeta$ &  $D_1$     \\
  2.5   &   0.5  &  4.4  &  0.8  &   0.43  & $-0.07$    \\ \hline
  $D_2$ & $D_3$ & $(pd\sigma)$ & $(pd\pi)$ & $(pp\sigma)$ & $(pp\pi)$ \\
   2.6  &  2.0  &  $-1.69$     &   0.78    &     0.55     &  $-0.14$  
\end{tabular}
\end{ruledtabular}
\label{tab:para}
\end{table}

\subsection{Previous views and \textit{ab initio} studies}

There are several views on the electronic structure of Ba$_2$IrO$_4$ in
the literature. Frequently it is being assumed that $e_g$ orbitals are 
empty in Ba$_2$IrO$_4$. In the present paper we make an attempt to find 
the electron densities in $e_g$ orbitals and in $2p$ oxygen orbitals. 
We are motivated by the \textit{ab initio} computations performed on a 
small cluster with embedding \cite{Sto12}. According to Mullikan 
population analysis the $5d$-shell charge on Ir ions is about $6.5e$, 
the effective ionic charge within $2p$ orbitals is about $5.2e$ on 
in-plane oxygens and about $5.6e$ on apical oxygens \cite{Court}. Note 
however that the direct mapping of these entries to the $d-p$ model can 
not be perfect as \textit{ab initio} computations include in addition 
oxygen valence $s$-orbitals, and also $s$, $p$ and $f$-orbitals at 
iridium ions (absent in the $d-p$ model).

Nonetheless, it seems clear that the formal idealized ionic model with
6 electrons occupying $2p$ levels of each oxygen and 5 electrons 
occupying $5d$ levels of each Ir ion can serve only as a rather crude 
starting view for the electronic structure of Ba$_2$IrO$_4$. For the 
sake of convenience let us introduce the notion of \textit{self-doping} 
$x$ for a single IrO$_4$ unit: with respect to the idealized formal 
ionic model where we have $x=0$; instead for the real compound we shall 
consider finite self-doping value $x=1.0$, i.e., one electron less per 
IrO$_4$ unit (according to Katukuri \cite{Sto12} $x\sim 0.8$; however 
we shall use $x=1$ instead because our cluster is rather small and 
while the self-doping $x=1$ translates well into integer total electron 
number in the cluster, other fractional values of $x$ would not.)

\section{The unrestricted Hartree-Fock approximation}
\label{sec:hf}

\subsection{The self-consistent Hartree-Fock problem}

We use the unrestricted HF approximation to investigate the
IrO$_4$ cluster (with cyclic boundary conditions). The technical 
implementation is the same as described in Refs. 
\cite{Miz96a,Miz01,rosc15}. Namely, the local Coulomb interaction 
Hamiltonian $H_{\rm int}$ is replaced by mean field terms derived in 
HF approximation. The averages
$\langle d^\dagger_{i\mu,\sigma}d_{i\nu,\sigma'}^{}\rangle$ 
(which appear within this treatment) can be treated as order parameters 
(there is a similar set of order parameters for oxygens). For the 
numerical calculations some initial values (some educated guess) have 
to be assigned to them to start the search for a self-consistent 
solution of the HF equations. During HF iterations the order parameters 
are recalculated self-consistently until convergence.

\subsection{Hartree-Fock calculations}

\begin{figure}[t!]
\begin{center}
\includegraphics[width=8.4cm]{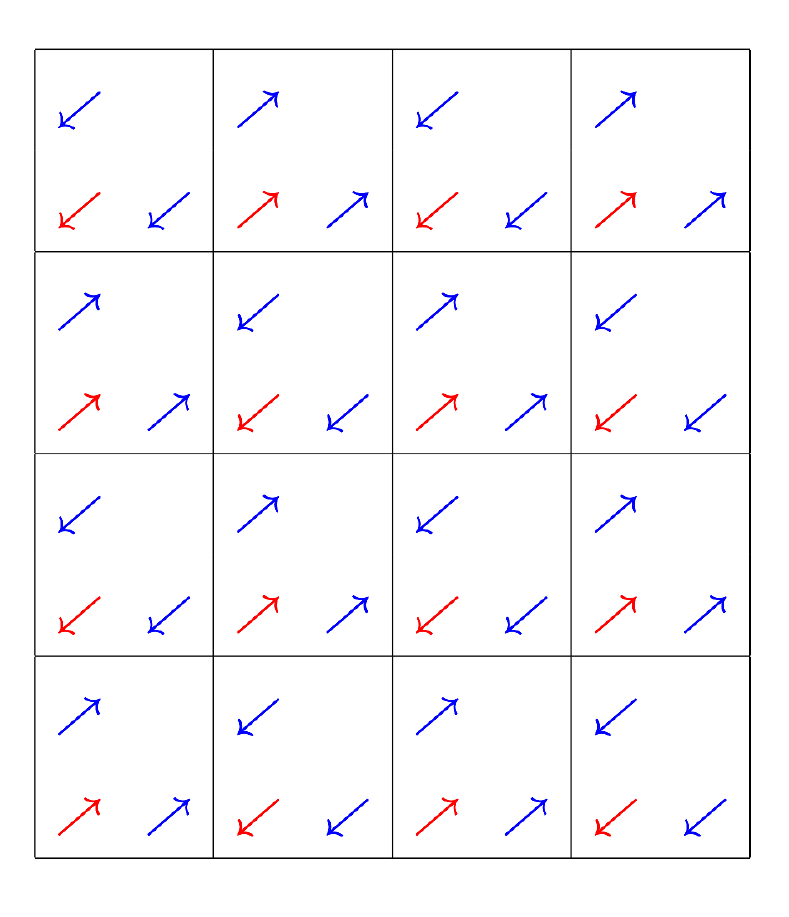}
\end{center}
\caption{(Color online) 
Antiferromagnetic spin alignment along (1,1,0) direction in  
a $4\times 4$ IrO$_4$ cluster. The arrows represent magnetic moments 
within an IrO$_2$ plane with the structure of CuO$_2$ plane in cuprates 
(apical oxygens are not shown), with red arrows for iridium ions, and 
blue arrows for in-plane oxygen ions; the spin magnitudes are 
arbitrary and only the directions are relevant. 
Vertical and horizontal lines are only guides to the eye --- they 
indicate IrO$_4$ units for a better visibility of individual 
symmetry and unit cells within the cluster.
}
\label{fig1}
\end{figure}

We are interested in charge-homogeneous solutions, in particular
homogeneity concerns primary order parameters,
$\langle d^\dagger_{i\mu,\sigma}d_{i\nu,\sigma}^{}\rangle$.
with $\mu = \nu$, i.e., the electron densities (occupations). To 
obtain unbiased results, we have studied several different types 
of order, and compared the energies of the self-consistently found
solutions to establish the ground state. Consequently, during the 
computations the following scenarios were studied:
\begin{itemize}
\item nonmagnetic phase, the $x,y,z$ spin components at all atoms 
      were set to be zero;
\item ferromagnetic (FM) phase with spins aligned along (1,1,0) direction;
\item FM phase with spins aligned along (1,0,0) direction 
      (parallel to the $a$ axis);
\item FM phase with spins (at all atoms) aligned along 
      (0,0,1) direction (parallel to the $c$ axis);
\item AF phase with spins (at all atoms) aligned along (1,1,0) 
      direction according to the pattern shown in Fig. \ref{fig1};
\item AF phase with spins (at all atoms) aligned along (1,0,0) direction 
      (like in Fig. \ref{fig1} but with spins rotated by 45 degrees);
\item AF phase with spins (at all atoms) aligned along 
      (0,0,1) direction.
\item in addition, all the magnetic phases were studied again with an 
      additional constraint that the magnetic moments at oxygen ions 
      vanish, i.e., for nonmagnetic oxygens.
\end{itemize}
Let us immediately comment on the the last scenario (magnetic iridiums 
and nonmagnetic oxygens): surprisingly, it yields too high HF energies 
and therefore it is never realized in the ground state. Therefore, we 
conclude that also oxygen ions contribute to the magnetic order by 
double exchange mechanism. 

\begin{table*}[t!]
\caption{Results of the HF calculations for the $d-p$ model 
(\ref{model}) concerning the magnetic order and degeneracy of HF ground 
state for: ionic model without self-doping ($x=0$) (top), and realistic 
model at self-doping $x=1.0$ (bottom), as obtained for increasing 
charge-transfer gap, $\Delta$ (\ref{Delta}). The quantities presented 
are: 
HOMO-LUMO gap $G$, 
the average energy of spin-orbit term (\ref{so-part}) per a single 
iridium ion $\langle SO\rangle$,
HF energy per one IrO$_4$ unit $E_{\rm HF}$. 
The obtained states are:  
AF1 --- AF state with moment alignment along (1,1,0) 
direction; 
AF2 --- AF state with moment alignment along (1,0,0) 
direction, i.e., along the $a$ axis; 
FM1 and FM2 are FM states with the numbers 1 and 2 having 
the same meaning like for AF states.  
The numbers in brackets (last column) are in (meV) and denote the 
energetic distance to the true (best) HF ground state.
Note that some low-energy states with a different magnetic order could 
not be detected during computations 
(denoted by ? in the last column). Note also that all these states are 
well insulating due to sizable HOMO-LUMO gaps $G$.
In contrast, the ground states obtained for the realistic model 
($x=1.0$) have very small or small HOMO-LUMO gaps $G$, and are weakly 
insulating or even close to conducting.
}
\begin{ruledtabular}
\begin{tabular}{cccccrcc}
  model &  $x$  &  $\Delta$(eV)  & G (eV)   &  $\langle SO\rangle$ (eV)  &  $E_{\rm HF}$ (eV)
        &  ground state & excited states \\ \hline
%\multicolumn{6}{c}{ $\mbox{idealistic ionic model for self-doping\;\;\;}  x=0$}   \\
%
ionic model& 0.0 & 1.0 &   0.12  &  $-0.36$  &  141.675  &   AF1  &   ?    \\
       &     & 1.5    &   0.29   &  $-0.41$  &  130.964  &   AF2  & AF1 (2); FM1 (2); FM2 (2) \\
       &     & 2.0    &   0.36   &  $-0.43$  &  120.093  &   AF1  & AF2 (0); FM1 (2); FM2 (2) \\
       &     & 2.5    &   0.41   &  $-0.46$  &  109.159  &   AF1  & AF2 (0); FM1 (0); FM2 (0) \\
       &     & 3.0    &   0.45   &  $-0.49$  &   98.162  &   AF1  & FM1 (0); AF2 (1); ?     \\
       &     & 3.5    &   0.46   &  $-0.54$  &   87.094  &   AF1  & FM1 (4); FM2 (3); ?      \\
       &     & 4.0    &   0.40   &  $-0.59$  &   75.952  &   AF1  & AF2 (5); FM1 (5); FM2 (5) \\
       &     & 4.5    &    0.31  &  $-0.64$  &   64.731  &   AF1  & FM1 (3); FM2 (3); AF2 (5) \\
       &     & 5.0    &    0.29  &  $-0.68$  &   53.436  &   AF1  & FM1 (0); FM2 (7); AF2 (7) \\  
       \hline
realistic model&1.0& 1.0 & 0.028  & $-0.41$  &  128.974  &   AF1  & AF2 (0); FM1 (1)    \\
       &     & 1.5    &   0.032   & $-0.43$  &  118.613  &   AF1  & AF2 (2);  ?  \\
       &     & 2.0    &   0.042   & $-0.45$  &  108.197  &   AF1  & AF2 (0); FM1 (0); FM2 (0) \\
       &     & 2.5    &   0.064   & $-0.48$  &  97.723   &   AF1  & FM1 (30); ?  \\
       &     & 3.0    &   0.154   & $-0.58$  &   87.134  &   AF1  & FM1 (1); ?        \\
       &     & 3.5    &   0.091   & $-0.61$  &   76.444  &   AF1  & FM1 (0); ?     \\
       &     & 4.0    &   0.019   & $-0.62$  &   65.663  &   FM1  & AF1 (3); ? \\
       &     & 4.5   &    0.129   & $-0.67$  &   54.846  &   AF2  & FM2 (0); AF1 (2) \\
       &     & 5.0   &    0.231   & $-0.68$  &   43.852  &   AF2  & FM1 (4.5); AF1 (4.5) \\
\end{tabular}
\end{ruledtabular}
\label{tab:ground-x0}
\end{table*}

The total number of different order parameters is large and results in 
a rather slow convergence of the HF procedure. The standard remedy for 
poor convergence is so-called dumping technique or a more sophisticated 
(possibly better) quantum chemistry technique called level shifting 
\cite{Sou73} (more details on the level-shifting technique as applied 
to the $d-p$ multiband model may be found in Ref. \cite{rosc15}).  
We have used both techniques in the present calculations. 
HF convergence criteria were the following conditions fulfilled 
simultaneously for the two consecutive iterations: 
(i) relative energy change should be smaller than $0.2\times 10^{-6}$; 
(ii) absolute charge change should be smaller than $0.5\times 10^{-3}e$; 
(iii) absolute magnetization change (per ion) should be smaller than 
$0.5\times 10^{-3}$.
Performing computations with dumping or with different level shifts,  also 
with different starting conditions (initial charge and magnetic 
densities) we completed many runs for each set of Hamiltonian 
parameters. The comparisons done afterwards lead us to believe that a  
practical accuracy of our HF solutions (on convergence) is about 1-2 meV.

\section{Numerical results}
\label{sec:resu}

Numerical studies of the multiband $d-p$ model (\ref{model}) require
not only the parameters which were fixed in Sec. \ref{sec:para}, but
also an assumption concerning the total electron number per unit cell.
We consider below two different scenarios:
(i) the idealistic formal ionic model with $n_0=5+4\times 6=29$ 
electrons per IrO$_4$ unit, and
(ii) the realistic model with a smaller total number of $n=29-x$ 
electrons, where we consider only one \textit{self-doping} namely
$x=1.0$ \cite{Sto12}.
Thereby we concentrate on the most important results obtained for 
realistic values of Coulomb interactions within the framework of 
these two different scenarios.

\subsection{Formal ionic model (self-doping $x=0$)}
\label{sec:ionic}

Taking the formal ionic model as a starting configuration for the HF
iterations, we assume that each IrO$_4$ unit has a negative charge 
$Q=29 e$ (the two Ba$^{2+}$ ions are considered only as donors of $4e$ 
to the IrO$_4$ unit). The initial charge distribution assumed on start 
of HF iterations quickly redistributes to reach the uniform and stable  
distribution during the iteration process (this is achieved due to 
finite $d-p$ hybridization).
The emerging ground states are shown in Table II:
for the ionic model with $x=0$ in the top part, and 
for the realistic model with self-doping $x=1$ (for the discussion 
see the next subsection) in the bottom part. 
From the data in the top part of Table II it follows that strong 
magnetic degeneracy occurs for the resulting magnetic ground states 
obtained in HF approximation. It is remarkable that the values of  
charge densities as well as average magnetic moments and average 
angular momenta are almost the same for degenerate FM and AF HF states. 
On the one hand, the experimentally observed antiferromagnet with 
moments aligned along (1,1,0) direction is confirmed, but on the other 
hand its energy is degenerated with other magnetic phases. 

\begin{figure}[t!]
\begin{center}
\includegraphics[width=8cm]{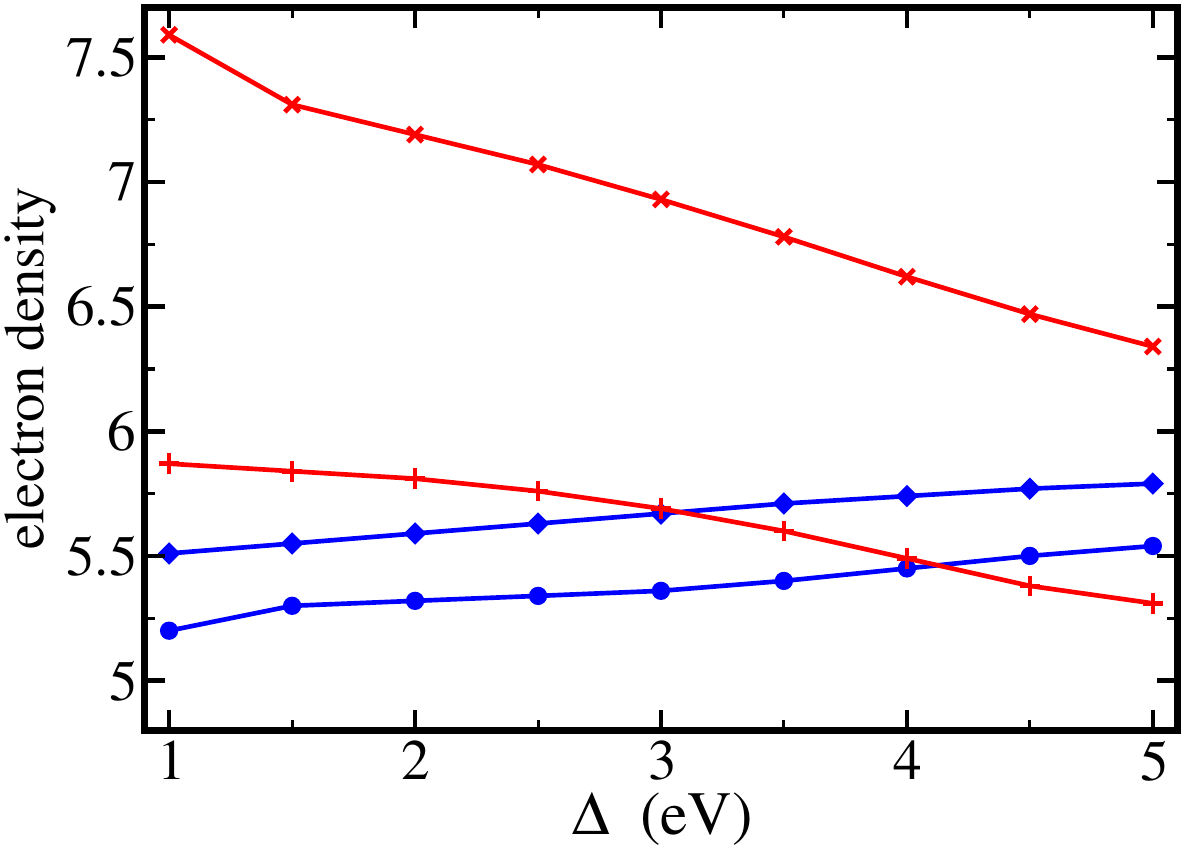}
\vskip .5cm
\includegraphics[width=8.1cm]{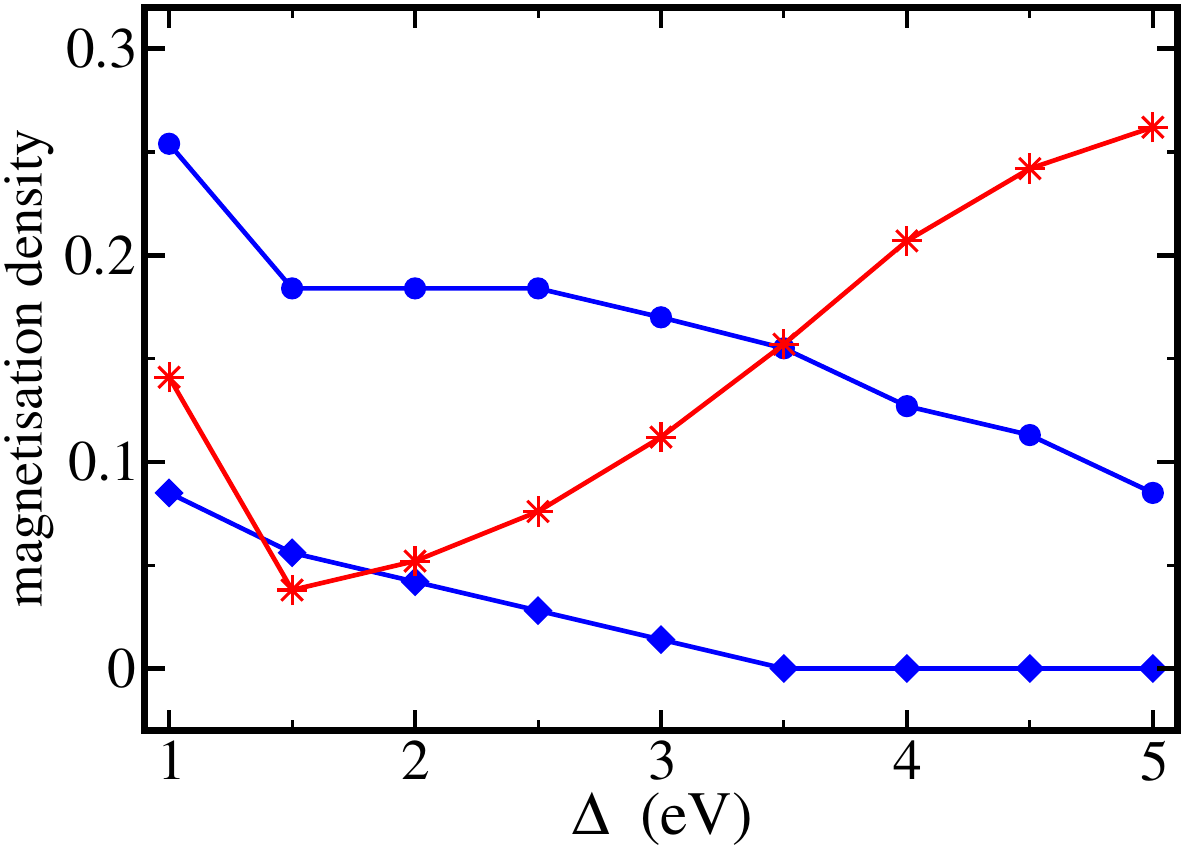}
\end{center}
\caption{(Color online) Electron occupation and spins in the multiband 
model for increasing $\Delta\equiv-\varepsilon_p$ obtained in the 
formal ionic model without self-doping ($x=0$). The investigated ground 
state is insulating with AF order aligned along (1,1,0) direction.
No significant orbital order was detected. 
Legend: data for Ir ions are represented by red lines; data for O ions 
are represented by blue lines;
data points show electron occupation numbers in
$t_{2g}$ orbitals (pluses),
all $5d$ orbitals ($\times$),
$2p$ orbitals at in-plane oxygens (circles), and
$2p$ orbitals at apical oxygens (diamonds).
}
\label{fig2}
\end{figure}

Finally, we present charge and magnetization densities obtained for the 
ionic model, see Fig. \ref{fig2}. As expected, increasing $\Delta$ 
reduces the charge density within $5d$ Ir orbitals due to weaker 
hybridization. Notably, there is a considerable electron transfer to 
$e_g$ orbitals which contain typically more than one $e_g$ electron per 
Ir ion. More precisely, the $e_g$ electron density decreases with 
increasing $\Delta$ from $\sim 1.7$ to $\sim 1.0$ in the investigated 
range of $\Delta$. Also charge transfer between $5d$ and $2p$ orbitals 
is found for increasing $\Delta$, and the charge densities increase for 
both nonequivalent oxygen positions, in-plane and apical. The density 
is closer to $\langle n_p\rangle=6$ obtained in the ionic picture for 
apical positions, where the hybridization plays a minor role. However, 
the most important feature is the observed disagreement of the results 
of Fig. \ref{fig2}. with the available \textit{ab initio} data 
\cite{Sto12}. The charge occupations and the magnetisation densities 
for AF (aligned along (1,1,0) direction) HF ground state are shown in 
Fig. \ref{fig2}.

\begin{figure}[t!]
\begin{center}
\includegraphics[width=8cm]{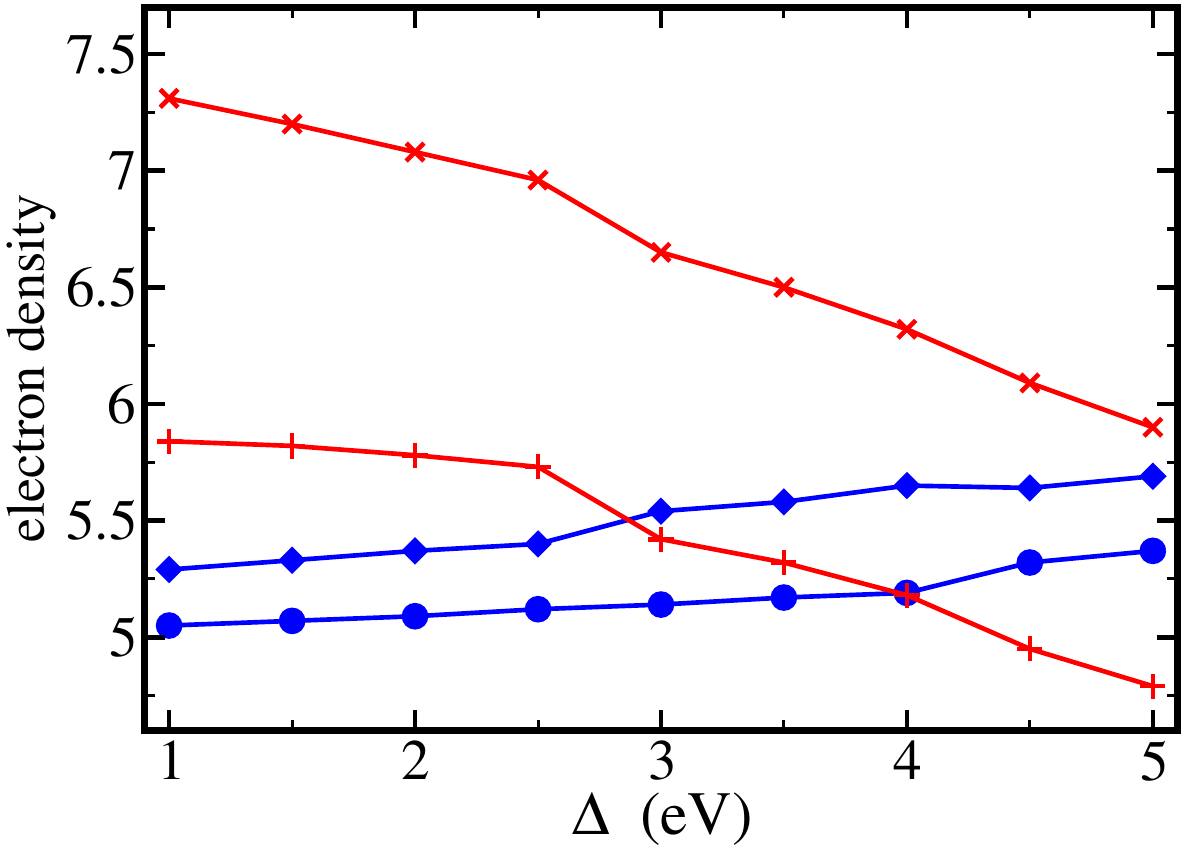}
\vskip .5cm
\includegraphics[width=8.1cm]{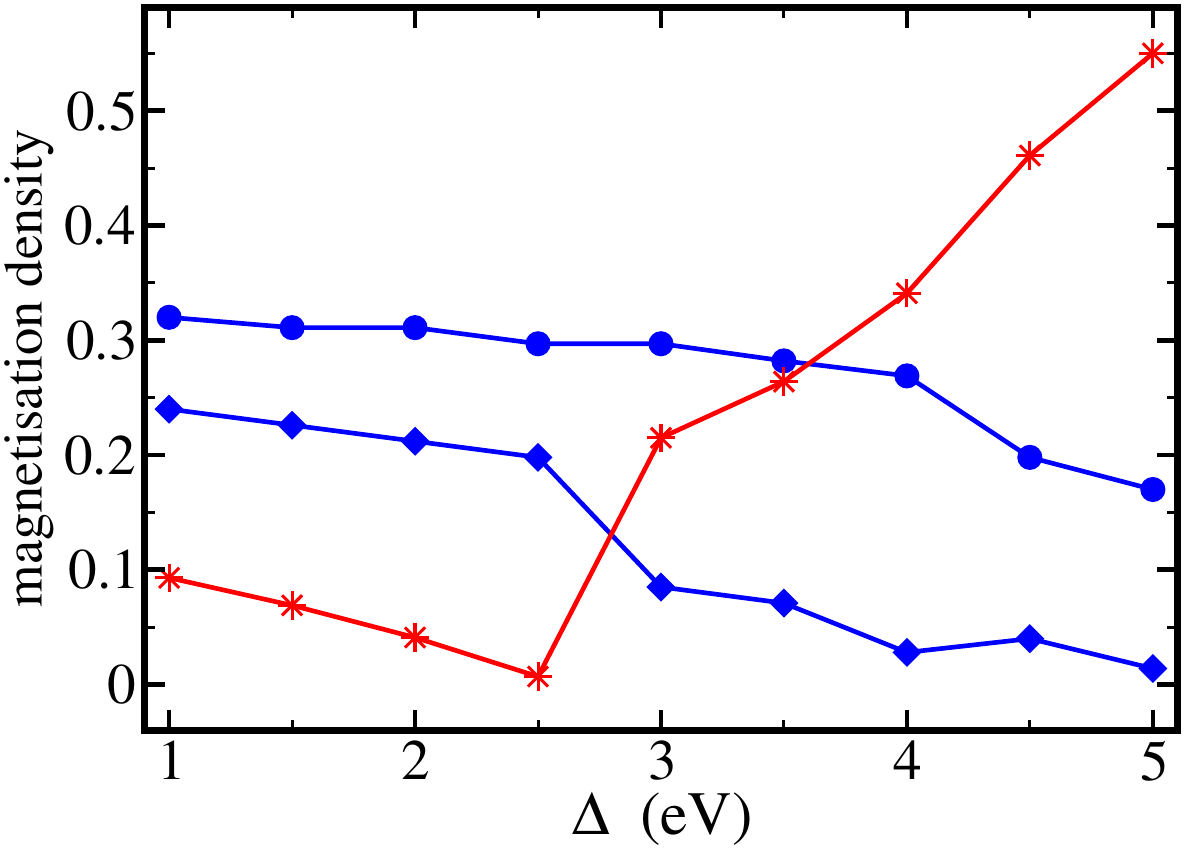}
\end{center}
\caption{(Color online) Electron occupations and magnetic moments in 
the multiband model for increasing $\Delta\equiv-\varepsilon_p$, 
obtained in the realistic model with self-doping $x=1$. 
The investigated ground state is weakly insulating with AF order 
aligned along (1,1,0) direction. No significant orbital order was 
detected. 
Legend: data for Ir ions are represented by red lines; data for O 
ions are represented by blue lines;
data points show electron occupation numbers in
$t_{2g}$ orbitals (pluses),
all $5d$ orbitals ($\times$),
$2p$ orbitals at in-plane oxygens (circles), and
$2p$ orbitals at apical oxygens (diamonds).
}
\label{fig3}
\end{figure}

\subsection{Realistic model with self-doping $x=1.0$}
\label{sec:self}

In the second scenario we follow the \textit{ab initio} results of 
Katukuri \textit{et al.} \cite{Sto12} and we assume a reduced total 
number of electrons per IrO$_4$ unit. Taking the total electron number 
$n=29-x$ with $x=1.0$ this corresponds to \textit{self-doping} by one 
hole. The obtained ground states and their characteristics are shown 
in the bottom part of Table II while the corresponding charge and 
magnetization densities are displayed in Fig. \ref{fig3}. Altogether,
taking a representative value of $\Delta=3.0$ eV one fnds enhanced 
electron density in $e_g$ orbitals and all magnetic moments larger by
a factor close to 2 than those obtained in the ionic model, cf. Figs. 
\ref{fig2} and \ref{fig3}. 

From the lower part of Table II we conclude that AF1 is the true ground 
state for a broad range of $\Delta\in [1.0,3.5]$ eV. For higher values
of $\Delta$ the identification of the true ground state is difficult 
due to strong degeneracy of the lowest energy states obtained from HF 
iterations for various magnetization distributions. Moreover, very 
small HOMO-LUMO gaps obtained frequently are direct evidence that
multiconfiguration HF computations would be necessary to establish the 
ground state and its magnetic order in a reliable way. However, such  
multiconfiguration HF computations are prohibitively costly for the 
considered cluster, both in terms of computer power and computation 
time. Therefore we present rather qualitative evidence which suggests 
that the magnetic order obtained in this range of $\Delta$ may compete 
with other magnetic states.

Thus both for idealistic ionic ($x=0$) and for realistic model with 
$x=1$ it is difficult to reach definite conclusions about the nature 
of the true ground state. However one bright point are computations 
for $x=1$ and $\Delta=3.0$ eV. Here the HOMO-LUMO gap $G$ of 0.154 eV 
is large enough to believe that HF (without correlations) performs 
satisfactorily. Moreover, this value is close to that reported in Ref. 
\cite{Mos14}, and the electron occupation numbers are quite close to 
the values reported in \textit{ab initio} calculations \cite{Sto12}. 
The magnetization magnitude at iridium ions is also roughly that which
was found in the experiment \cite{isobe12}.

\subsection{Sensitivity of the results to variation of Hubbard repulsion $U_d$}

Consider first the electron distribution obtained for the self-doping 
model, see Table III. First of all, one finds almost one hole per ion
for O$_1$ oxygen postions in the IrO$_2$ planes, and roughly half 
of the hole at apical O$_3$ positions. This demonstrates a considerable 
hole delocalization over oxygen orbitals. Furthermore, also $e_g$ 
orbitals are partly filled, as one finds the electron numbers: 
$n_{t2g}=5.42$ and $n_{eg}=1.24$ per one Ir ion. Large electron density 
in $e_g$ orbitals may be explained by a stronger $\sigma$-bond between 
$x^2-y^2$ orbitals and O($2p_{\sigma}$) orbitals within the IrO$_2$ 
planes. These numbers suggest that even in the effective models which 
feature only Ir($5d$) orbitals also $e_g$ orbitals have to be included. 

\begin{table}[t!]
\caption{ 
Average electron densities $\langle n\rangle$, average magnetization
densities $\langle S^{\alpha}\rangle$, and average orbital momenta 
$\langle L^{\alpha}\rangle$, with $\alpha=x,y,z$, as obtained for 
$x=1$ and $\Delta=3.0$ eV for AF1 ground state. Average charge 
densities are close to those reported by \textit{an-initio} studies 
\cite{Sto12}. HOMO-LUMO gap $G=0.154$ eV (weak insulator). Altogether 
the obtained data agree quite well with the experimental results.
Legend: O$_1$ --- in plane oxygen; O$_3$ --- apical oxygen; Ir --- 
iridium ion.
Note that for AF1 state the signs of the $\langle S^{\alpha}\rangle$
and $\langle L^{\alpha}\rangle$ components alternate
between sublattices.
}
\begin{ruledtabular}
\begin{tabular}{ccclcrrc}
 atom & $\langle n\rangle$ & $\langle S^x\rangle$ & $\langle S^y\rangle$ & $\langle S^z\rangle$ 
                           & $\langle L^x\rangle$ & $\langle L^y\rangle$ & $\langle L^z\rangle$ \\ 
\hline
  Ir   & 6.65 & 0.152 & 0.152 &  0.0  & $-0.18$ &  $0.22$ &  0.0  \\
 O$_1$ & 5.14 & 0.21  & 0.21  &  0.0  &  $0.03$ &  $0.03$ &  0.0  \\
 O$_3$ & 5.54 & 0.06  & 0.06  &  0.0  & $-0.01$ & $-0.01$ &  0.0  \\
\end{tabular}
\end{ruledtabular}
\label{tab:orderpar-x1}
\end{table}

The order parameters presented in Table III suggest that physical
description of the IrO$_4$ plane in the framework of the realistic 
model (with self-doping $x=1$), with charge occupations close to the 
\textit{ab initio} data, and with Hamiltonian parameters from Table I
fits the experimental data reasonably well. However, a natural question 
arises: Was this achieved by accident or are the obtained results 
generic (at least in some limited sense) ? To get a better insight into 
the problem we performed numerous additional 
computations for different sets of Hamiltonian parameters (we could not 
afford the full study of the phase diagram as the number of parameters 
involved and already discussed technical issues make the computations 
rather expensive).

The results of these computations indicate that the obtained results 
are indeed generic and the selection of the parameters is quite 
representative. To give an explicit example let us consider two
additional sets of Hamiltonian parameters:
the first set with  $U_d=2.$0 eV and $J_d=0.4$ eV with 
weaker correlations at iridium ions, and the second set with stronger 
correlations for $U_d=3.0$ eV and $J_d=0.6$ eV. All other Hamiltonian 
parameters remain unchanged (note that we keep the ratio $J_d/U_d=0.2$).
We make scans varying $\Delta$ for realistic scenario when self-doping 
$x=1$ is selected. We made a search for and accepted results which give 
the occupation numbers close to \textit{ab initio} results, i.e., the 
$n_{t2g}+n_{eg}$ electron occupation on Ir close to 6.5. The collected 
results are presented in Table IV (compare with Table III).

\begin{table}[t!]
\caption{ 
Average electron densities $\langle n\rangle$, average magnetization
densities $\langle S^{\alpha}\rangle$, and average orbital momenta 
$\langle L^{\alpha}\rangle$, with $\alpha=x,y,z$, as obtained for 
different atoms at self-doping $x=1$ and for two different sets of 
Hamiltonian parameters:\\
--- Top part --- for large $\Delta=5.0$ eV, $E_{\rm HF}=38.7174$ eV, 
HOMO-LUMO gap $G=0.138$ eV; the values of order parameters are 
almost the same for both (energy degenerate) FM and AF ground states 
(for AF1 phase they alternate between the two sublattices); parameters: 
$U_d=2.0$ eV, $J_d=0.4$ eV.\\
--- Lower part --- for small $\Delta=1.5$ eV, $E_{\rm HF}=124.8861$ eV, 
HOMO-LUMO gap $G=0.124$ eV; the ground state is FM (FM1) but the energy 
of nearly degenerate AF1 state is only by 50 meV higher; parameters: 
$U_d=3.0$ eV, $J_d=0.6$ eV.
}
\begin{ruledtabular}
\begin{tabular}{cccccrrc}
 atom & $\langle n\rangle$ & $\langle S^x\rangle$ & $\langle S^y\rangle$ & $\langle S^z\rangle$ 
                           & $\langle L^x\rangle$ & $\langle L^y\rangle$ & $\langle L^z\rangle$ \\ 
\hline
\multicolumn{8}{c}{ AF1 ground state  } \\
\hline
  Ir   & 6.59 & 0.162 & 0.162 &  0.0  & $-0.20$ & $-0.22$ &  0.0  \\
 O$_1$ & 5.16 & 0.20  & 0.20  &  0.0  &  $0.03$ &  $0.03$ &  0.0  \\
 O$_3$ & 5.54 & 0.06  & 0.06  &  0.0  &  $0.0$  &  $0.0$  &  0.0  \\
\hline\hline
\multicolumn{8}{c}{  FM1 ground state  } \\
\hline
  Ir   & 6.58 & 0.169 & 0.169 &  0.0  & $-0.20$ & $-0.25$ &  0.0  \\
 O$_1$ & 5.18 & 0.19  & 0.19  &  0.0  &  $0.03$ &  $0.03$ &  0.0  \\
 O$_3$ & 5.53 & 0.07  & 0.07  &  0.0  &  $0.0$  &  $0.0$  &  0.0  \\
\end{tabular}
\end{ruledtabular}
\label{tab:orderpar2-x1}
\end{table}

We emphasize that not only the electron densities and and magnetic
moments are similar for the two sets of parameters used in Table IV, 
but also the populations of $t_{g}$ and $e_g$ orbitals at It ions.
For the first choice with $U_d=2.0$ eV, one finds: $n_{t2g}=5.38$ and
$n_{eg}=1.22$. In the case of larger $U_d=3.0$ eV, one finds: 
$n_{t2g}=5.40$ and $n_{eg}=1.18$. Note that these densities are also 
quite close to the ones obtained for the parameters given in Table I.

To summarize, we remark that FM and AF ground states with magnetic 
moments aligned along (1,1,0) direction are very close to each other. 
Nonmagnetic ground state and states with the moments aligned along 
(0,0,1) direction are much higher in energy, thus they can be safely 
excluded.

\subsection{Can magnetic moments in IrO$_4$ plane be aligned 
along $z$ axis?  }

The next question one should try to answer is the magnetic anisotropy.
Why almost all magnetic states are aligned parallel to the plane, and 
what about alignments (if any) along the $z$ axis? In fact, there are 
also ground states with magnetic moments aligned along the $z$ axis 
(perpendicular to the iridiums-oxygen plane). They occur both for the 
idealistic ionic model, (i.e., for $x=0$) and for the realistic model 
(with self-doping $x=1$). The conditions for them to appear are the 
following: large $U_d$ and large $\Delta$ (and what follows large 
charge occupations on oxygens, close to 6$e$). 
Here we also present an example in Table V.

The electron densities obtained for large values of $\Delta$ and the 
magnetic states with magnetic oriented along the $z$ spin axis show
a much better hole localization at Ir ions. The hole densities at O$_1$
and O$_3$ oxygen positions are reduced by more than a half with respect 
to their values obtained for the states presented in Tables III and IV.
With more holes within $5d$ orbitals, also electron densities are 
reduced to $n_{t2g}=4.41$ and $n_{eg}=1.00$, but again the density of 
$e_g$ electrons is large and also in this case it is necessary to 
include all five $5d$ orbitals at Ir ions.  

\begin{table}[t!]
\caption{ 
Average charge occupations $\langle n\rangle$, average magnetic spin 
components $\langle S^{\alpha}\rangle$, and average orbital momenta 
$\langle L^{\alpha}\rangle$, with $\alpha=x,y,z$, obtained for different 
ions the AF ground state with magnetic moments \textit{aligned along 
the $z$ axis}. Note that the energy of this AF state is degenerate with 
that of FM state (with moments also aligned along the $z$ axis).
The Hamiltonian parameters are: $U_d=3.0$ eV, $J_d=0.6$ eV,  
$\Delta=5.0$ eV, and $x=1.0$. 
}
\begin{ruledtabular}
\begin{tabular}{cccccccc}
 atom & $\langle n\rangle$ & $\langle S^x\rangle$ & $\langle S^y\rangle$ & $\langle S^z\rangle$ 
                           & $\langle L^x\rangle$ & $\langle L^y\rangle$ & $\langle L^z\rangle$ \\ 
\hline
  Ir   & 5.41 & 0.0 & 0.0 &  0.78  & 0.0 & 0.0 &  0.43  \\
 O$_1$ & 5.52 & 0.0 & 0.0 &  0.01  & 0.0 & 0.0 &  0.04  \\
 O$_3$ & 5.72 & 0.0 & 0.0 &  0.00  & 0.0 & 0.0 &  0.03  \\
\end{tabular}
\end{ruledtabular}
\label{tab:orderpar3}
\end{table}

It might be tempting to speculate that for a ground state 
antiferromagnetism aligned along (1,1,0) or (1,0,0) direction results 
from the optimization of kinetic hoppings parallel to the 
iridium-oxygen plane and this happens only for substantial hole 
occupations on oxygen orbitals. On the other hand, when charge densities 
within $2p$  oxygen orbitals become much closer to six (no holes), this 
blocks effective long-range hopping and only then the possibility of 
magnetization along the z axis may become an option.

\subsection{Do correlations favor antiferromagnetic over ferromagnetic order ?}

The energy-degenerate ground states: FM and AF (see Table II) are a 
puzzle. The question arises which one is the true ground state. 
On the HF level this can not be answered but only after one includes 
electronic correlation one can get the final answer. In general the 
task to compute the correlations is very difficult and costly. There 
is however a simple (though not rigorous) intuitive approach which 
suggests that the AF order should be favored in the ground state.

Namely let us consider configuration-interaction computations used 
in quantum chemistry to derive the correlation energy, as defined for 
instance in Refs. \cite{Fulde,Levine}. Let us limit ourselves only 
to double-excitions from the occupied to virtual states. For the 
Hamiltonian (\ref{model}) with the parameters from Table I 
(i.e., with large $U_d$ in the interaction part of the Hamiltonian) 
it is expected that singlet excitations are dominant and the triplet 
excitions are probably much less important.

Now consider intraatomic correlations (each of the double excitions is
coming out from occupied levels on some single ion \cite{Fulde}). These 
are strictly local, i.e., for AF or FM ground state they look the same 
and most probably give the same contribution to the total correlation 
energy. (Thus they do not differentiate between AF and FM order).

In contrast, interatomic correlations only involve excitions from 
\textit{different} pairs of ions \cite{Fulde}. The consequence is that, 
naively speaking, the number of the occupied electron pairs of the 
singlet-type (and located on different ions) is much bigger for AF 
ground state. Each such occupied singlet-pair (in AF) during CI 
computations is annihilated (while the virtual singlet-pair is created 
at the same time) giving substantial contribution to the total singlet 
correlation energy. The number of occupied triplet-pairs of the 
electrons is smaller and the triplet excitions give much smaller 
correlation contribution anyway (we remind that for the Hamiltonian 
(\ref{model}) singlet excitations are probably dominant).

For the FM ground state we can apply a similar reasoning. Now, the 
number of occupied triplet pairs of the electrons (on different ions) 
is bigger. However, the annihilation of each such pair gives rise to 
a much smaller triplet contribution to the total correlation energy.

In summary, one can expect that the total correlations contribution
to the total ground state energy is mainly of the singlet-type
and that the correlations contribution to the total energy is
enhanced for AF ground state while being weaker for FM ground state.
Thus one expects that AF order should win in the ground state.

\section{Discussion and summary}
\label{sec:summa}

We have shown that the HF computational results for the assumed 
idealized ionic configuration do not describe properly the 
experimental data for Ba$_2$IrO$_4$ for almost all studied values of 
$\Delta$. Namely, the charge on iridium ions is 7$e$-7.5$e$, i.e., it
is significantly higher than 6.5$e$ (as suggested by \textit{ab initio}
\cite{Sto12}) and the HOMO-LUMO gap $G$ is also too large, so the 
system is a good insulator. Only for very large $\Delta=5.0$ eV the 
agreement with \textit{ab initio} calculations (charges on iridium 
and also on oxygens) improves but still the HOMO-LUMO gap is twice 
larger than 0.14 eV expected from the experimental data 
\cite{Oka11,Mos14}. Note that large insulating gap was found instead 
in Na$_2$IrO$_3$ \cite{Com12}.

In contrast, the realistic model with self-doping $x=1$ is doing much
better. The charges on iridium are generally smaller
(in comparison to the $x=0$ case) and in  particular for $\Delta=3.0$ 
eV the charge occupations on both iridium and oxygens are very close 
to those indicated by \textit{ab initio} \cite{Sto12} and the ground 
state:  
(i) is antiferromagnetic aligned along (1,1,0);
(ii) is weakly insulating (not far from being metal; $G=0.15$ eV, 
i.e., the value indicated by experimental data \cite{Oka11,Mos14}); 
(iii) has magnetic moments with magnitude being close to the 
experimental findings \cite{Oka11,Mos14}; 
(iv) is characterized by some traces of weak orbital order.

We emphasize that the present treatment of the spin-orbit interaction
is sufficient to break the SU(2) symmetry in the spin space and to 
realize the experimentally observed type of antiferromagnetic phase. 
Further justification to the self-doping concept comes from the hole
delocalization over oxygen orbitals, found practically for all 
considered parameter sets. This delocalization implies that also Ba
ions cannot be considered as Ba$^{2+}$ which would be the case of the 
idealistic ionic model. Furthermore, using the analogy to cuprates, we 
suggest that also CuO$_2$ planes in the undoped high-$T_c$ materials 
\cite{Arr09,Ole87,Lor02,Med09} cannot be considered as charged 
formally in the same way as predicted in the ionic model, i.e., 
Cu$^{2+}$O$_2^{2-}$, but in agreement with recent results \cite{Gas14}, 
self-doping is also here an important effect to include.

Notably, in the Hartree-Fock analysis quantum fluctuations are 
neglected and thus both ferromagnetic and antiferromagnetic phases are 
energetically degenerate. But by a simple reasoning one can come to 
conclusion that when including these effects and correlations beyond 
Hartree-Fock, the antiferromagnetic order observed in experiment 
should become more stable.

There are other important messages. Namely when describing IrO$_4$ 
plane in the framework of $d-p$ model numerous interesting physical 
facts and electronic mechanisms become visible. In this respect the 
$d-p$ model is a very good supplement to simpler but more widely used 
five-band model (featuring only Ir ions with five $5d$-type occupied 
Wannier orbitals which are coming from hybridization of iridium $5d$ 
and oxygen $2p$ orbitals). Our results serve as a guideline that 
effective models such as the five-band model have to be constructed 
including their proper electron densities and thus indicate the 
importance of self-doping. In other words, the formal ionic picture 
(with zero self-doping) most probably does not describe faithfully the 
electronic states, and to introduce self-doping is a must. 
Unfortunately, in order to do it properly some preliminary 
\textit{ ab initio} investigations (in a small cluster) are necessary.

After completing this paper, we became aware of a very recent paper 
which presents a complete study of the superexchange model \cite{Sol15}. 
The AF phase obtained by us from the multiband model captures the 
interplay between the spin-orbit coupling and charge excitations in 
a charge-transfer insulator and agrees well with the one 
which follows from the low-energy superexchange model.   

\acknowledgments

We are grateful to V. M. Katukuri for providing us with the results 
of Mulliken population analysis which we have used in this paper.
We kindly acknowledge support by Narodowe Centrum Nauki
(NCN, National Science Center) under Project No. 2012/04/A/ST3/00331.


\begin{thebibliography}{99}


\bibitem{Moo09} S. J. Moon, H. Jin, W. S. Choi, J. S. Lee, S. S. A. Seo, 
                   J. Yu, G. Cao, T. W. Noh, and Y. S. Lee,
                   Phys. Rev. B \textbf{80}, 195110 (2009).
 
\bibitem{Car91} S. A. Carter, J. Yang, T. F. Rosenbaum, J. Spa\l{}ek, 
                   and J. M. Honig,
                   Phys. Rev. B \textbf{43}, 607 (1991);
                X. Yao, J.~M.~Honig, T. Hogan, C. Kannewurf, and J. Spa\l{}ek,
                   \textit{ibid.} \textbf{54}, 17469 (1996).
 
\bibitem{Ari12} R. Arita, J. Kune\v{s}, A. V. Kozhevnikov, A. G. Eguiluz, 
                   and M. Imada,
                   Phys. Rev. Lett. \textbf{108}, 086403 (2012);
                R.~Arita, J. Kune\v{s}, P. Augustinsky, A. V. Kozhevnikov,
                   A.~G.~Eguiluz, and M. Imada,
                   J. Phys. Soc. Conf. Proc. \textbf{3}, 013023 (2014).

\bibitem{li13}  Q. Li, G. Cao, S. Okamoto, J. Yi, W. Lin, B. C. Sales, J. Yan,
                   R. Arita, J. Kune\v{s}, A. V. Kozhevnikov, A.~G.~Eguiliz,
                   M. Imada, Z. Gai, M. Pan, and D.~G.~Mandrus,
                   Sci. Rep. \textbf{3}, 3073 (2013).

\bibitem{Hsi12} D. Hsieh, F. Mahmood, D. H. Torchinsky, G. Cao, and N. Gedik,
                   Phys. Rev. B \textbf{86}, 035128 (2012).

\bibitem{Moo06} S. J. Moon, M. W. Kim, K. W. Kim, Y. S. Lee, J.-Y.~Kim,
                   J.-H. Park, B. J. Kim, S.-J. Oh, S. Nakatsuji, Y.~Maeno,
                   I. Nagai, S. I. Ikeda, G. Cao, and T.~W.~Noh,
                   Phys. Rev. B \textbf{74}, 113104 (2006).

\bibitem{Wat10} H. Watanabe, T. Shirakawa, and S. Yunoki,
                   Phys. Rev. Lett. \textbf{105}, 216410 (2010).

\bibitem{kim12} B. H. Kim, G. Khaliullin, and B. I. Min,
                   Phys. Rev. Lett. \textbf{109}, 167205 (2012).

\bibitem{Kimx2} J. Kim, D. Casa, M. H. Upton, T. Gog, Y.-J. Kim, 
                   J.~F.~Mitchell, M. van Veenendaal, M. Daghofer, 
                   J.~van den Brink, G. Khaliullin, and B. J. Kim,
                   Phys. Rev. Lett. \textbf{108}, 177003 (2012).
 
\bibitem{Wat13} H. Watanabe, T. Shirakawa, and S. Yunoki,
                   Phys. Rev. Lett. \textbf{110}, 027002 (2013).

\bibitem{Ham15} A. Hampel, C. Piefke, and F. Lechermann,  
                   Phys. Rev. B \textbf{92}, 085141 (2015).

\bibitem{Oka11} H. Okabe, M. Isobe, E. Takayama-Muromachi, A.~Koda, 
                   S. Takeshita, M. Hiraishi, M. Miyazaki, R.~Kadono, 
                   Y. Miyake, and J. Akimitsu,
                   Phys. Rev. B \textbf{83}, 155118 (2011).

\bibitem{Mos14} S. Moser, L. Moreschini, A. Ebrahimi, B.~Dalla Piazza,
                   M. Isobe, H. Okabe, J. Akimitsu, V.~V.~Mazurenko,
                   K.~S.~Kim, A. Bostwick, E. Rotenberg, J.~Chang,
                   H.~M.~R\o{}nnow, and M. Grioni,
                   New J. Phys. \textbf{16}, 013008 (2014).

\bibitem{Tor15} A. dela Torre, S. McKeown Walker, F. Y. Bruno, S.~Ricc\'o, 
                   Z. Wang, I. Gutierrez Lezama, G. Scheerer, G.~Giriat, 
                   D.~Jaccard, C. Berthod, T. K. Kim, M.~Hoesch, 
                   E.~C.~Hunter, R. S. Perry, A. Tamai, and F. Baumberger,  
                   Phys. Rev. Lett. \textbf{115}, 176402 (2015).

\bibitem{Kim08} B. J. Kim, H. Jin, S. J. Moon, J.-Y. Kim, B.-G. Park, 
                   C.~S.~Leem, J. Yu, T. W. Noh, C. Kim, S.-J. Oh, 
                   J.-H.~Park, V. Durairaj, G. Cao, and E. Rotenberg,
                   Phys. Rev. Lett. \textbf{101}, 076402 (2008).
                   
\bibitem{Jac09} G. Jackeli and G. Khaliullin,
                   Phys. Rev. Lett. \textbf{102}, 017205 (2009);
                G. Khaliullin, 
                   \textit{ibid.}   \textbf{111}, 197201 (2013);
                J.~Chaloupka and G. Khaliullin,   
                   \textit{ibid.}   \textbf{116}, 017203 (2016).

\bibitem{Brz15} W. Brzezicki, A. M. Ole\'s, and M. Cuoco,
                   Phys. Rev. X \textbf{5}, 011037 (2015);
                W. Brzezicki, M. Cuoco, and A.~M.~Ole\'s,
                   J. Sup. Novel Magn. \textbf{29}, in press (2016).

\bibitem{Bos13} S. Boseggia, R. Springell, H. C. Walker, H. M. Ronnow,
                   Ch. R\"uegg, H. Okabe, M. Isobe, R. S. Perry,
                   S.~P.~Collins, and D. F. McMorrow,
                   Phys. Rev. Lett. \textbf{110}, 117207 (2013);
                S. Boseggia, \textit{Magnetic order and excitations in
                   perovskite iridates studied with resonant X-ray
                   scattering techniques}, PhDwhat is thesis
                   (Univ. College London, London, 2014).

\bibitem{Sal14} M. Moretti Sala, M. Rossi, S. Boseggia, J. Akimitsu,
                   N. B. Brookes, M. Isobe, M. Minola, H. Okabe, 
                   H.~M.~R\o{}nnow, L. Simonelli, D. F. McMorrow, and 
                   G.~Monaco,
                   Phys. Rev. B \textbf{89}, 121101(R) (2014);
                M.~Moretti Sala, M. Rossi, A. Al-Zein, S. Boseggia,
                   E.~C.~Hunter, R.~S.~Perry, D. Prabhakaran, A. T. Boothroyd,
                   N.~B.~Brookes, D.~F.~McMorrow, G. Monaco, and M.~Krisch,
                   \textit{ibid.} \textbf{90}, 085126 (2014).

\bibitem{Kat14} V. M. Katukuri, V. Yushankhai, L. Siurakshina, 
                   J. van den Brink, L. Hozoi, and I. Rousochatzakis,
                   Phys. Rev. X \textbf{4}, 021051 (2014).
                   
\bibitem{Tro10} F. Trousselet, A. M. Ole\'s, and P. Horsch,
                   Europhys. Lett. \textbf{91},  40005 (2010);
                   Phys. Rev. B    \textbf{86}, 134412 (2012).
                   
\bibitem{Cha10} J. Chaloupka, G. Jackeli, and G. Khaliullin,  
                   Phys. Rev. Lett. \textbf{105}, 027204 (2010);
                                    \textbf{110}, 097204 (2013);
                F. Trousselet, M. Berciu, A.~M.~Ole\'s, and P. Horsch,
                   \textit{ibid.}   \textbf{111}, 037205 (2013);
                J. G. Rau, E. K.-H. Lee, and H.-Y. Kee,
                   \textit{ibid.}   \textbf{112}, 077204 (2014);
                K. Hu, Fa Wang, and Ji Feng,
                   \textit{ibid.}   \textbf{115}, 167204 (2015).
                   
\bibitem{Uch14} M. Uchida, Y. F. Nie, P. D. C. King, C. H. Kim,
                   C.~J.~Fennie, D. G. Schlom, and K. M. Shen,
                   Phys. Rev. B \textbf{90}, 075142 (2014).

\bibitem{Sto12} V. M. Katukuri, H. Stoll, J. van den Brink, and L. Hozoi,
                   Phys. Rev. B \textbf{85}, 220402(R) (2012);
                V. M. Katukuri, \textit{Quantum chemical approach to
                   spin-orbit excitations and magnetic interactions in
                   iridium oxides},
                   PhD Thesis (Technische Universit\"{a}t Dresden, 2014).

\bibitem{rosc15} K. Ro\'sciszewski and A. M. Ole\'s,
                   Phys. Rev. B \textbf{91}, 155137 (2015).

\bibitem{Miz96a} T. Mizokawa and A. Fujimori,
                   Phys. Rev. B \textbf{54} 5368 (1996).

\bibitem{Pol12} L. V. Poluyanov, and W. Domcke,
                   J. Chem. Phys. \textbf{137}, 114101 (2012);
                H. Matsuura and K. Miyake,
                   J. Phys. Soc. Jpn. \textbf{82}, 073703 (2013);
                L. Du, L. Huang, and X. Dai,
                   Eur. Phys. J. B \textbf{86}, 94 (2013).
                   
\bibitem{oguchi95} T. Oguchi, 
                   Phys. Rev. B \textbf{51}, 1385 (1995);
                C. Noce and M. Cuoco,
                   \textit{ibid.} \textbf{59}, 2659 (1999).

\bibitem{Gri71} J. S. Griffith,
                     \textit{The Theory on Transition Metal Ions}
                     (Cambridge University Press, 1971).

\bibitem{Ole05} A. M. Ole\'s, G. Khaliullin, P. Horsch, and L. F. Feiner,
                   Phys. Rev.  B \textbf{72}, 214431 (2005).

\bibitem{Hor07} P. Horsch,
                   \textit{Orbital Physics in Transition-Metal Oxides:
                   Magnetism and Optics}, in: Handbook of Magnetism and
                   Advanced Magnetic Materials,
                   edited by H. Kronm\"uller, and S. Parkin,
                   Volume 1: Fundamentals and Theory (J.~Wiley and Sons, Ltd., 2007).

\bibitem{Ole12} A. M. Ole\'s and G. Stollhoff,
                   Phys. Rev.  B \textbf{29}, 314 (1984).
                   
\bibitem{Har05} W. A. Harrison,
                   \textit{Elementary Electronic Structure},
                   (World Scientific, London, 2005).

\bibitem{Maz12} I. I. Mazin, H. O. Jeschke, K. Foyevtsova, R. Valent\'i, 
                   and D.~I.~Khomskii,
                   Phys. Rev. Lett. \textbf{109}, 197201 (2012).
% for na2iro3  U 1-2 eV   Jh ~0.5 eV

\bibitem{Com12} R. Comin, G. Levy, B. Ludbrook, Z.-H. Zhu, C.~N.~Veenstra,
                   J. A. Rosen, Yogesh Singh, P. Gegenwart, D.~Stricker,
                   J. N. Hancock, D. van der Marel, I.~S.~Elfimov,
                   and A. Damascelli,
                   Phys. Rev. Lett. \textbf{109}, 266406 (2012).

\bibitem{Li15}  Y. Li, K. Foyevtsova, H. O. Jeschke, and R. Valent\'i,
                   Phys. Rev. B \textbf{91}, 161101(R) (2015).

\bibitem{Yam14} Y. Yamaji, Y. Nomura, M. Kurita, R. Arita, and M.~Imada,
                   Phys. Rev. Lett. \textbf{113}, 107201 (2014).
 
\bibitem{hunds} D. van der Marel and G. A. Sawatzky,
                   Phys. Rev. B \textbf{37}, 10674 (1988);
                A. Georges, L. de' Medici, and J. Mravlje,
                   Annu. Rev. Condens. Matter Phys. \textbf{4}, 137 (2013).

\bibitem{Kho03} D. I. Khomskii and M. V. Mostovoy,
                     J. Phys. A: Math. Gen. \textbf{36}, 9197 (2003).

\bibitem{Arr09} E. Arrigoni, M. Aichhorn, M. Daghofer, and W. Hanke,
                     New J. Phys. \textbf{11}, 055066 (2009).

\bibitem{Hyb89} M. S. Hybertsen, M. Schl\"uter, and N. E. Christensen,
                   Phys. Rev. B \textbf{39}, 9028 (1989).

\bibitem{Gra92} J. B. Grant and A. K. McMahan,
                   Phys. Rev. B \textbf{46}, 8440 (1992).

\bibitem{Esk91} H. Eskes and G. A. Sawatzky,
                   Phys. Rev. B \textbf{44}, 9656 (1991).

\bibitem{Shi08} A. Shitade, H. Katsura, J. Kune\v{s}, X.-L. Qi, 
                   S.-C. Zhang, and N. Nagaosa,
                   Phys. Rev. Lett. \textbf{102}, 256403 (2009).

\bibitem{Court} The data obtained by courtesy of V. M. Katukuri.

\bibitem{Miz01} T. Mizokawa, L. H. Tjeng, G. A. Sawatzky, G. Ghiringhelli,
                   O. Tjernberg, N. B. Brookes, H. Fukazawa, S.~Nakatsuji,
                   and  Y. Maeno,
                   Phys. Rev. Lett. \textbf{87}, 077202 (2001).

\bibitem{Sou73} V. R. Sounders and I. H. Hillier,
                     Int. J. Quant. Chem. \textbf{7}, 699 (1973).

\bibitem{isobe12} M. Isobe, H. Okabe, E. Takayama-Muromachi, A. Koda,
                     S. Takeshita, M. Hiraishi, M. Miyazaki, R. Kadono,
                     Y.~Miyake, and J. Akimitsu,
                     J. Phys.: Conf. Series, \textbf{400}, 032028 
                                                       and 032071 (2012).

\bibitem{Fulde} P. Fulde, \textit{Electron Correlations in Molecules and Solids},
                   Springer Series in Solid-State Sciences,
                   Vol. \textbf{100} (Springer-Verlag, Berlin, 1995).
 
\bibitem{Levine} I. N. Levine,
                   \textit{Quantum Chemistry} (Prentice Hall, New Jersay, 2013).
                   
\bibitem{Ole87} A. M. Ole\'s, J. Zaanen, and P. Fulde,
                   Physica B\&C {\bf 148}, 260 (1987).
                   
\bibitem{Lor02} J. Lorenzana and G. Seibold,
                   Phys. Rev. Lett. \textbf{89}, 136401 (2002).
                     
\bibitem{Med09} L. de' Medici, X. Wang, M. Capone, and A. J. Millis,
                   Phys. Rev. B \textbf{80}, 054501 (2009);
                A. Thomson and S.~Sachdev,
                   \textit{ibid.} \textbf{91}, 115142 (2015).

\bibitem{Gas14} H. Ebrahimnejad, G. A. Sawatzky, and M. Berciu, 
                   Nature~Phys. \textbf{10}, 951 (2014).

\bibitem{Sol15} I. V. Solovyev, V. V. Mazurenko, and A. A. Katanin,
                   Phys. Rev. B \textbf{92}, 235109 (2015).

\end{thebibliography}
\end{document}